\documentstyle[twocolumn,prl,aps,epsf]{revtex}
\begin{document}
%\draft
%%%%%%%%%%%% Begin Cover Page %%%%%%%%%%%%%%%%%%%%%%%%%%%%%%%%%%%%%%%%%%
\preprint{\hfill ANL-HEP-PR-99-05}
\title{\vspace*{-1ex} \hfill {\small ANL-HEP-PR-99-05}
\vspace*{1ex} \\
Single-top-squark production via $R$-parity-violating supersymmetric
couplings \\
in hadron collisions}
\author{Edmond L. Berger, B. W. Harris, and Z. Sullivan} 
\address{High Energy Physics Division,
             Argonne National Laboratory,
             Argonne, Illinois 60439 }
\date{March 31, 1999}
\maketitle

\begin{abstract}
Single-top-squark production via $q q^{\prime} \rightarrow
\overline{\widetilde{t}}_1$ probes $R$-parity-violating extensions of
the minimal supersymmetric standard model though the
$\lambda^{\prime\prime}_{3ij}$ couplings.  For masses in the range
180--325 GeV, and $\lambda^{\prime\prime}_{3ij} >$ 0.02--0.06, we show
that discovery of the top squark is possible with 2 fb$^{-1}$ of
integrated luminosity at run II of the Fermilab Tevatron.  The bound
on $\lambda^{\prime\prime}_{3ij}$ can be reduced by up to an order of
magnitude with existing data from run I, and by two orders of
magnitude at run II if the top squark is not found.
\end{abstract}

\vspace*{5ex}
%\vspace*{0.2in}
%\pacs{PACS numbers: 11.30.Pb, 12.60.Jv, 14.80.Ly}
%%%%%%%%%%%% End of Cover Page %%%%%%%%%%%%%%%%%%%%%%%%%%%%%%%%%%%%%%%%%

In supersymmetric extensions of the standard model, particles may be
assigned a new quantum number called $R$-parity ($R_p$)~\cite{farrar}.
The particles of the standard model are $R_p$ even, and their
corresponding superpartners are $R_p$ odd.  The bounds on possible
$R_p$-violating couplings are relatively restrictive for the first two
generations of quarks and leptons, but much less so for states of the
third generation~\cite{dreiner}.  If $R_p$ is conserved, as is often
assumed, superpartners must be produced in pairs, each of which decays
to a final state that includes at least one stable lightest
supersymmetric particle (LSP).  The production rates for pairs of
strongly interacting supersymmetric particles, the squarks and
gluinos, benefit from the large color couplings of these superpartners
to the incident light quarks and gluons in hadronic scattering
subprocesses.  However, in many models, the squarks and gluinos are
relatively heavy, and therefore their pair production incurs a large
phase space suppression.

In this Letter, the $s$-channel production of a {\it{single}} squark
through an $R_p$-violating mechanism~\cite{savas} is considered.  The
motivation is that the greater phase space offsets the reduced
coupling strength in the production.  The focus is on the relatively
light top squark $\widetilde{t}_1$ and its subsequent $R_p$-{\it
conserving} decays.  Thus, the $R$-parity violation penalty is paid only
once, in the initial production, and is offset by the greater phase
space relative to pair production.

Beginning with the superpotential for $R_p$-violating couplings, we
write the partonic cross section for the process $q q^{\prime}
\rightarrow \overline{\widetilde{t}}_1$ and compare with that for pair
production.  Then discussing observability, we focus on one clean
$R_p$-conserving decay, $\widetilde{t}_1 \rightarrow b
\widetilde{\chi}^+_1$, with $\widetilde {\chi}^+_1 \rightarrow l^+ +
\nu +\widetilde {\chi}^0_1$.  Here, $l$ is an electron or muon, and
the $\widetilde {\chi}^+_1$ and $\widetilde {\chi}^0_1$ are the
chargino and lowest-mass neutralino states of the minimal
supersymmetric standard model (MSSM).  For top-squark masses in the
range of 180--325 GeV, we simulate both the signal and standard model
background processes and thereby show that the top squark can be
discovered, or the current bound on the size of the $R_p$-violating
couplings $\lambda^{\prime\prime}_{3ij}$ can be reduced by up to one
order of magnitude with existing data and by two orders of magnitude
at the forthcoming run~II of the Fermilab Tevatron.

In general it is possible to have $R_p$-violating contributions to the
MSSM superpotential of the baryon- or lepton-number violating
type. However, limits on the proton decay rate severely restrict their
simultaneous presence.  We therefore assume the existence of a
baryon-number-violating coupling only of the form~\cite{weinberg}
\begin{equation}
{\cal W}_{\not \! R_p} = \lambda_{ijk}^{\prime\prime}U_i^cD_j^cD_k^c \; .
\end{equation}
Here, $U^c_i$ and $D^c_i$ are right-handed-quark singlet chiral
superfields, $i,j,k$ are generation indices, and $c$ denotes charge
conjugation.

In four-component Dirac notation, the Lagrangian that follows from this 
superpotential term is
\begin{eqnarray}
{\cal L}_{\lambda^{\prime\prime}} & = & -2 \epsilon^{\alpha \beta \gamma}
\lambda^{\prime\prime}_{ijk}
\left[ \widetilde u_{R i \alpha} \overline{d^c_{j \beta}} P_R d_{k \gamma}
  +    \widetilde d_{R j \beta} \overline{u^c_{i \alpha}} P_R d_{k \gamma}
\right. \nonumber \\
     && \hspace*{6em} + \left. 
     \widetilde d_{R k \gamma} \overline{u^c_{i \alpha}} P_R d_{j \beta} 
\right] + h.c. \; ,
\end{eqnarray}
where $j < k$.  For production of a right-handed top squark via an
$s$-channel diagram $\bar d^j\bar d^k \rightarrow \widetilde{u}^i_R $
the relevant couplings are $\lambda^{\prime\prime}_{312}$,
$\lambda^{\prime\prime}_{313}$, and $\lambda^{\prime\prime}_{323}$.
The most direct limits on these couplings come from the measurement of
$R_l$, the partial decay width to hadrons over the partial decay width
to leptons of the $Z$ boson.  For the top-squark masses considered,
$R_l$ provides 95\% confidence-level upper bounds of
$\lambda^{\prime\prime}_{3ij} < 1$ \cite{dreiner}.

The color- and spin-averaged partonic cross section for
inclusive $\widetilde{t}_1$ production is
\begin{equation}
\hat{\sigma} =
\frac{2\pi}{3} |\lambda^{\prime\prime}_{3ij}|^2
\frac{\sin^2\theta_{\widetilde{t}}}{m^2_{\widetilde{t}_1}} \delta(1 -
m^2_{\widetilde{t}_1}/\hat{s}) \; ,
\end{equation}
where $\sqrt{\hat{s}}$ is the partonic center of mass energy, and
$\theta_{\widetilde{t}}$ relates the right- and left-handed top-squark
interaction states to the mass eigenstates.  The hadronic cross
section depends on the following combinations of incident parton
distribution functions (PDF's): $d \otimes s$, $d \otimes b$, and $s
\otimes b$, where $d$, $s$, and $b$ denote the PDF's of the down,
strange, and bottom quarks, respectively.

\begin{figure}[tb]
\begin{center}
\epsfxsize= 3in %actual
\leavevmode
\epsfbox{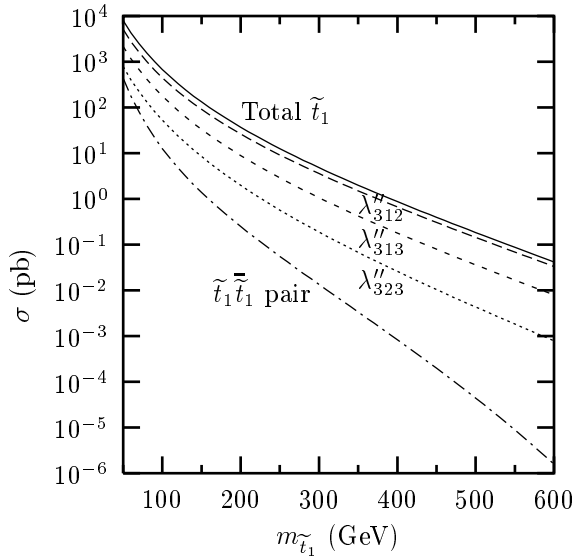}
\end{center}
\caption{Cross section for $R$-parity-violating production of a single
top squark at run II of the Fermilab Tevatron ($\sqrt{S}=2$ TeV) with
$\lambda^{\prime\prime}_{3ij}=0.1$ compared with the
$R$-parity-conserving production cross section for top-squark pairs
versus $m_{\widetilde{t}_1}$.}
\label{fig1}
\end{figure}

The mass dependences of the cross sections for single and
pair~\cite{zerwas} production of top squarks differ significantly, as
shown in Fig.~\ref{fig1}.  The curves are based on CTEQ4L PDF's
\cite{CTEQ4} and $\lambda^{\prime\prime}_{3ij} = 0.1$.  Even if
$\lambda^{\prime\prime}_{3ij}$ is reduced to $0.01$, two orders of
magnitude below the current bound, the single-top-squark rate exceeds
the pair rate for all $m_{\widetilde{t}_1}>100$ GeV.  The parton
luminosities determine that the contribution to the total cross
section of the terms proportional to $\lambda^{\prime\prime}_{312}:
\lambda^{\prime\prime}_{313}: \lambda^{\prime\prime}_{323}$ is about
$0.75:0.20:0.05$ at $m_{\widetilde{t}_1}=200$ GeV.  For simplicity, we
define $\lambda^{\prime\prime} \equiv \lambda^{\prime\prime}_{312} =
\lambda^{\prime\prime}_{313} = \lambda^{\prime\prime}_{323}$.  Our
numerical results represent the sum of $\widetilde{t}_1$ and
$\overline{\widetilde{t}}_1$ production.

An evaluation of the possibility for detection of a single top squark
requires discussion of the likely decay modes of the squark and an
estimation of standard model backgrounds.  In the $R_p$-conserving
MSSM, the up-type squark $\widetilde u^k$ can decay into charginos
and neutralinos via the two-body processes $\widetilde u^k\rightarrow
d^k+\widetilde{\chi}^+_j$ ($j=1,2$) and $\widetilde u^k\rightarrow
u^k+\widetilde \chi^{0}_j$ ($j=1,2,3,4$), where $\widetilde \chi^+_j$
and $\widetilde{\chi}^0_j$ represent a chargino and neutralino,
respectively.  Various three body modes are possible, including
$\widetilde t_1 \rightarrow W^+ + b +\widetilde \chi_1^{0}$, which is
similar to decay into the top quark (followed by top decay to $W^+ +
b$) but softer; and $\widetilde t_1 \rightarrow c + \widetilde
\chi^{0}_j$ via a flavor-changing loop process.

In the $R_p$-violating MSSM, the right-handed up-type squark
$\widetilde u^k_R$ can also decay into quark pairs $\widetilde
u_R^k\rightarrow \bar d^j +\bar d^i$ via the $\lambda^{\prime\prime}$
couplings.  The branching fraction into two jets is shown in 
Fig.~\ref{fig2}.  If $\lambda^{\prime\prime}$ is large, the decay to
quark jets dominates.  However, as shown below, the
$R_p$-conserving decay still produces a measurable and useful 
cross section.
  
\begin{figure}[tb]
\begin{center}
\epsfxsize= 3in  %actual
\leavevmode
\epsfbox{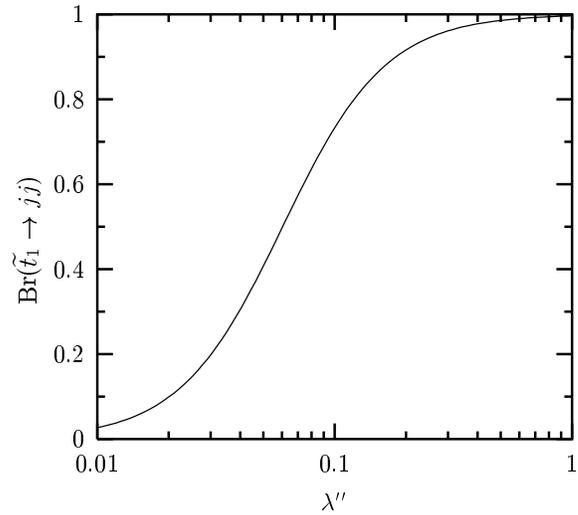}
\end{center}
\caption{Branching ratio for the top squark to decay into two jets
via the $R$-parity-violating coupling $\lambda^{\prime\prime}$
as a function of the coupling.}
\label{fig2}
\end{figure}

For the remainder of this Letter, we focus on the two-body decay
mode~\cite{dreinerross} $\widetilde{t}_1 \rightarrow b +\widetilde
{\chi}^+_1$, with $\widetilde \chi^{+} \rightarrow l + \nu +
\widetilde \chi_1^{0}$.  Here, $l$ denotes an electron or muon, which
usually comes from a $W$.  When $R$-parity is violated, $\widetilde
\chi_1^{0}$ is no longer stable; however, its lifetime is long
($c\tau > 100$ m, cf. the first paper of Ref.~\cite{dreiner}), and thus
we expect it to decay outside of the detector.

To obtain the relevant masses and decay branching fractions, we adopt
a minimal supergravity model~\cite{sugra}.  We begin with common
scalar and fermion masses of $m_0 = 100$ GeV and $m_{1/2} = 150$ GeV,
respectively, at the Grand Unified Theory (GUT) scale.  We choose a
trilinear coupling $A_0 = -300$ GeV and the ratio of the Higgs vacuum
expectation values $\tan\beta = 4$.  The absolute value of the Higgs
mass parameter $\mu$ is fixed by electroweak symmetry breaking and is
assumed positive.  Superpartner masses and decay widths are calculated
with ISAJET~\cite{Paige}.  At the weak scale, $m_{\widetilde{t}_1}=$
183 GeV, $m_{\widetilde {\chi}^0_1}=$ 55 GeV, $m_{\widetilde
{\chi}^{\pm}_1}=$ 103 GeV, and $\sin\theta_{\widetilde{t}} = 0.8$.  In
order to isolate the effects of the $R_p$-violating sector, we vary
$m_0$ and keep the other supersymmetric parameters fixed.  Since the
gaugino masses depend primarily on the choice of $m_{1/2}$, variation
of $m_0$ allows us to vary $m_{\widetilde t_1}$ without any
appreciable change in the masses of the decay products, or the mixing
angle $\theta_{\widetilde{t}}$.

The signal of interest consists of a tagged $b$-quark jet, a lepton,
and missing transverse energy associated with the unobserved neutrino
and $\widetilde{\chi}_1^{0}$.  The dominant backgrounds, in order of
importance, arise from production and decay of the standard model
processes $Wc$, with a charm quark $c$ that is mistaken for a $b$;
$Wj$, with a hadronic jet that mimics a $b$; $Wb\bar b$; $Wc\bar c$;
and single-top-quark production via $Wg$ fusion.  For these background
processes, we work with tree-level matrix elements obtained from
MADGRAPH~\cite{madgr} convolved with leading-order CTEQ4L \cite{CTEQ4}
parton distribution functions, at a hard scattering scale $\mu^2 =
\hat{s}$.  In an experimental analysis, the $Wj$ background will be
normalized by the data.  To simulate the resolution of the hadron
calorimeter, we smear the jet energies with a Gaussian whose width is
$\Delta E_j/E_j = 0.80/\sqrt{E_j}\oplus 0.05$ (added in quadrature).

We simulate the acceptance of the detector by using the selections
listed in Table~\ref{acceptance}.  The assumed coverage in rapidity
for taggable $b$-quark jets and leptons is smaller for run~I than for
run~II.  However, the signal and background are similar in shape in
these variables, and thus $S/B$ is not sensitive to this cut.  The
lepton must be isolated from any jets, as defined by a cone of radius
$\Delta R$, or it is considered missed.  Similarly, an isolation cut
is used for the $b$-quark jet in order to help identify it.  We assume
a $b$-tagging efficiency of $60\%$ ($50\%$ for run~I) with a mistag
rate of $15\%$ for charm quarks and $0.5\%$ for light
quarks~\cite{Yao}.

As expected from the primary decay, $\widetilde{t}_1 \rightarrow b
+\widetilde{\chi}^+_1$, the distribution in the transverse energy
$E_T$ of the $b$ quark is peaked sharply near the maximum value
allowed kinematically.  The spectrum of the background $b$ quark is
soft, and thus we impose a hard cut ($E_{Tb}>40$ GeV) on the minimum
$E_T$ of the $b$-quark jet.  The $b$-jet becomes too soft to be
detected if $m_{\widetilde {t}_1} \le m_{\widetilde {\chi}_1} +
E_T^{\rm cut}$.  This contributes to a lower limit on $m_{\widetilde
{t}_1}$ below which our proposed search mode is not useful.

The background from single-top-{\it quark} production will produce a
peak in any mass reconstruction.  We utilize the fact that single
top quarks are often produced with extra hard jets, and impose a ``jet
veto''.  We require that there be no hard jets ($E_{Tj}>20$ GeV,
$|\eta|<2.5$), beyond the one that is $b$-tagged, in the hadron
calorimeter.  After the jet veto, the remaining background is due
almost entirely to misidentified charm and light-quark jets from $Wc$
and $Wj$ production.  The transverse energy of the lepton tends to be
relatively soft for the signal at lower $m_{\widetilde t_1}$, whereas
it peaks around 45~GeV when it comes from the $W$ in the background.
A cut to remove hard leptons, with $E_{Tl}>45$ GeV, reduces the
background by a factor of 2 with little effect on the signal at low
masses.  The final significance for the signal at run~II is barely
changed by this ``lepton veto'', but it is especially helpful for the
run~I data.
\begin{table}
\caption{Cuts used to simulate the acceptance of the detector at the
Tevatron run~II, and run~I (in parentheses if different).
The lepton veto ($E_{Tl}<45$ GeV) is optimized for small top-squark mass.
\label{acceptance}} 
\begin{center}
\begin{tabular}{ll}
$|\eta_b|<2$ (1) & $E_{Tb}>40$ GeV \\
$|\eta_l|<2.5$ (1.1) & $E_{Tl}>15$ GeV (20 GeV)\\
$|\eta_j|<2.5$ & $E_{Tj}>20$ GeV \\
$|\Delta R_{bj}|>0.7 $ & $|\Delta R_{jl}|>0.7$ \\
${\not \!E}_{T}>20$ GeV & $E_{Tl}<45$ GeV
\end{tabular} \end{center} 
\end{table}

\begin{figure}[tb]
\begin{center}
\epsfxsize= 3in  %actual
\leavevmode
\epsfbox{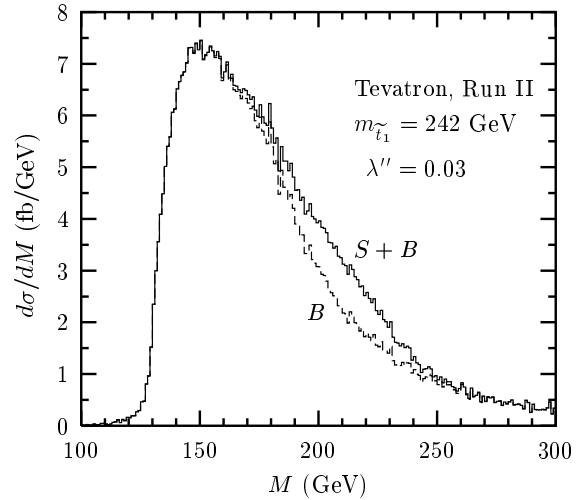}
\end{center}
\caption{The reconstructed-mass $M$ distribution for single-top-squark
production ($S$) and backgrounds ($B$) at the Tevatron ($\sqrt{S}=2$
TeV) for a top-squark mass $m_{\widetilde{t}_1}=242$ GeV.  The
coupling $\lambda^{\prime\prime}=0.03$ produces the minimum signal for
a $5\sigma$ significance at this mass.}
\label{fig3}
\end{figure}

Shown in Fig.~\ref{fig3} is an example of the signal and background.
For this case, $m_{\widetilde {t}_1} = 242$ GeV and
$\lambda^{\prime\prime} = 0.03$.  The mass variable is defined as
$M^2=(P_b + P_l + P_X)^2$ where the $P_b$ and $P_l$ are the
four-momenta of the $b$ and lepton.  The four-momentum $P_X$ is
defined such that its three-momentum balances that of the $b$ and
lepton, and $P_X^2 \equiv 0$.  The reconstructed $Wj$ background turns
on at $m_W + E_{Tb}^{\rm cut}$, and thus peaks at 150~GeV, before
falling rapidly with mass.  The signal in Fig.~\ref{fig3} would
constitute a discovery at the level of $5\sigma$ with an integrated
luminosity of 2~fb$^{-1}$ at $\sqrt{S} = 2$ TeV.  The significance is
calculated for a mass window of $\pm30$ GeV about the center of the
peak.  A change of the window size to either $\pm20$ GeV or $\pm40$
GeV produces the same significance to within a few percent.  When
$m_{\widetilde{t}_1}$ is reduced to 183 GeV, the signal and background
spectra peak at about the same location, and sensitivity to the signal
begins to be lost.

\begin{figure}[tb]
\begin{center}
\epsfxsize= 3in  %actual
\leavevmode
\epsfbox{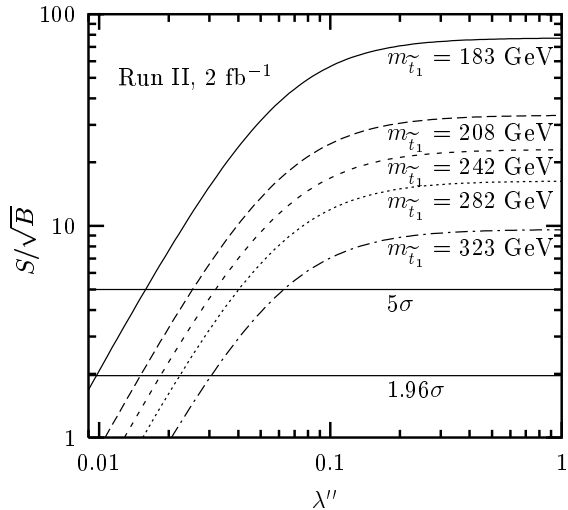}
\end{center}
\caption{Statistical significance of the single-top-squark signal
($S/\sqrt{B}$) in run~II of the Tevatron ($\sqrt{S}=2$ TeV, 2 fb$^{-1}$)
versus $\lambda^{\prime\prime}$ for a variety of top-squark masses.}
\label{fig4}
\end{figure}

Examination of the structure of the cross section involving $R_p$-{\it
conserving} decay modes reveals that as $\lambda^{\prime\prime}$
grows, the decrease in branching fraction is compensated by the
increase in cross section.  As depicted in Fig.~\ref{fig4},
\begin{equation}
\sigma \propto \frac{|\lambda^{\prime\prime}|^2} 
{|\lambda^{\prime\prime}|^2 + f(R_p)} \;,
\end{equation}
where $f(R_P)$ is a constant times the branching fraction into
$R_p$-conserving modes.  As $\lambda^{\prime\prime} \to \infty$, the
cross section goes to a constant; whereas, when
$\lambda^{\prime\prime} \to 0$ the cross section decreases as
$|\lambda^{\prime\prime}|^2$.  The relationship $S/\sqrt{B} \propto
{|\lambda^{\prime\prime}}|^2/\sqrt{B}$, valid for small
$\lambda^{\prime\prime}$, implies a lower limit on the values of
$\lambda^{\prime\prime}$ that can be probed.  On the other hand, this
relationship highlights an insensitivity to variations in the estimate
of the background.

\begin{figure}[tb]
\begin{center}
\epsfxsize= 3in  %actual
\leavevmode
\epsfbox{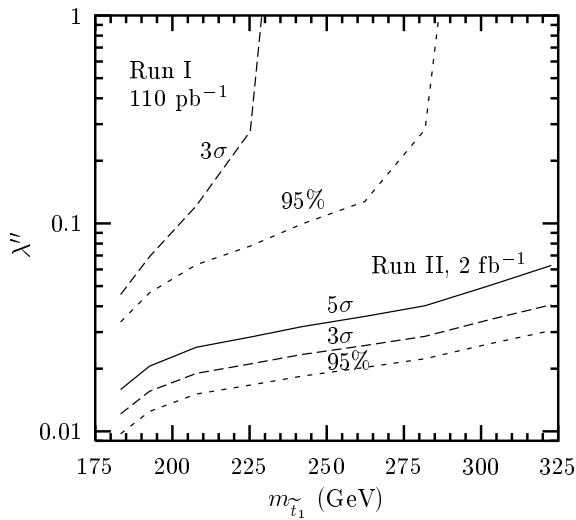}
\end{center}
\caption{Lower limits on discovery ($S/\sqrt{B} = 5$), evidence
($S/\sqrt{B} = 3$), and 95\% confidence-level exclusion ($S/\sqrt{B} =
1.96$) for $\lambda^{\prime\prime}$ versus top-squark mass in run~I of
the Tevatron ($\sqrt{S}=1.8$ TeV, 110~pb$^{-1}$), and in run~II
($\sqrt{S}=2$ TeV, 2~fb$^{-1}$).}
\label{fig5}
\end{figure}

In Fig.~\ref{fig5}, we show the reach in $\lambda^{\prime\prime}$ for
$180 < m_{\widetilde{t}_1} < 325$ GeV.  With an integrated luminosity
of 2~fb$^{-1}$ at $\sqrt{S} = 2$ TeV, discovery at the level of
$5\sigma$ is possible provided that $\lambda^{\prime\prime} >$
0.02--0.06.  Otherwise, a 95\% confidence-level exclusion can be set
for $\lambda^{\prime\prime} >$ 0.01--0.03. For the lower integrated
luminosity and energy of the existing run~I data, values of
$\lambda^{\prime\prime} >$ 0.03--0.2 can be excluded at the 95\%
confidence level if $m_{\widetilde{t}_1}=$180--280 GeV.  In the limit
that only one of the $\lambda^{\prime\prime}_{3ij}$ couplings is
non-zero, a conservative estimate of the sensitivity may be obtained
by dividing by the relative contribution of the coupling to the
production cross section presented earlier.

We conclude that, as long as the $R_p$-conserving decay $\widetilde
{t}_1 \rightarrow b \widetilde {\chi}^+_1 \rightarrow \l \nu
\widetilde{\chi}^0_1$ is allowed, it should be possible to discover
the top squark at run~II of the Fermilab Tevatron, for $180 <
m_{\widetilde{t}_1} < 325$ GeV and $\lambda^{\prime\prime}_{3ij} >$
0.02--0.06, or to lower the direct limit on $\lambda^{\prime\prime}$
by two orders of magnitude.  Existing data from run~I of the Tevatron
should allow a reduction of the limit on $\lambda^{\prime\prime}$ by
an order of magnitude.  With such a reduction, one can establish that
$R_p$-violating decay is unlikely and rule out most of the possible
influence of the top squark on single-top-quark production and decay.
These points and details of our calculation will be presented
elsewhere.

%%% Add other thanks here %%%
We thank Herbi Dreiner, Tao Han, Frank Paige, and Carlos Wagner.  This
work was supported by the U.S. Department of Energy, High Energy
Physics Division, under Contract No. W-31-109-Eng-38.

%%%%%%%%%%%%%% Begin References %%%%%%%%%%%%%%%%%%%%%%%%%%%%%%%%%%%%%%%%

%%%%%%%%%%%%%% End of References %%%%%%%%%%%%%%%%%%%%%%%%%%%%%%%%%%%%%%%

%%%%%%%%%%%%%% Begin Tables  %%%%%%%%%%%%%%%%%%%%%%%%%%%%%%%%%%%%%%%%%

%%%%%%%%%%%%%% End of Tables %%%%%%%%%%%%%%%%%%%%%%%%%%%%%%%%%%%%%%%%%

%%%%%%%%%%%%%% Begin Figures %%%%%%%%%%%%%%%%%%%%%%%%%%%%%%%%%%%

%%%%%%%%%%%%%% End of Figures %%%%%%%%%%%%%%%%%%%%%%%%%%%%%%%%%%%%%%%


\begin{references}

\bibitem{farrar}
 P.~Fayet, Nucl. Phys. {\bf B90}, 104 (1975); 
 A.~Salam and J.~Strathdee, Nucl. Phys. {\bf B87}, 85 (1975).

\bibitem{dreiner}
 B.~Allanach {\it et al.}, in {\it Proceedings of the Workshop on
Physics at Run II -- Supersymmetry/Higgs}, Fermilab, 1998 (to be published),
hep-ph/9906224;
 G.~Bhattacharyya, D.~Choudhury, and K.~Sridhar, Phys. Lett. B {\bf 355}, 
 193 (1995).

\bibitem{savas}
 S.~Dimopoulos, R.~Esmailzadeh, L.~J.~Hall, J.-P.~Merlo, and
 G.~D.~Starkman, Phys. Rev. D {\bf 41}, 2099 (1990).

\bibitem{weinberg}
 S.~Weinberg, Phys. Rev. D {\bf 26}, 287 (1982); 
 N.~Sakai and T.~Yanagida, Nucl. Phys. {\bf B197}, 533 (1982).

\bibitem{zerwas}
 W.~Beenakker, M.~Kr{\" a}mer, T.~Plehn, M.~Spira, and P.~M.~Zerwas, 
 Nucl.~Phys. {\bf B515} 3, (1998).

\bibitem{CTEQ4}
 CTEQ Collaboration, H.~Lai {\it et al.}, Phys. Rev. D {\bf 55}, 1280 (1997).

\bibitem{dreinerross}
 H.~Dreiner and G.~G.~Ross, Nucl.~Phys. {\bf B365}, 597 (1991).

\bibitem{sugra}
 A.~Chamseddine, R.~Arnowitt, and P.~Nath, Phys. Rev. Lett. {\bf 49}, 970 
 (1982); R.~Barbieri, S.~Ferrara, and C.~A.~Savoy, Phys. Lett. B{\bf 119}, 
 343 (1982); L.~J.~Hall, J.~Lykken, and S.~Weinberg, Phys. Rev. D {\bf 27}, 
 2359 (1983).

\bibitem{Paige} 
 F.~E.~Paige, S.~D.~Protopopescu, H.~Baer, and X.~Tata, 
 Brookhaven report BNL-HET-98-18, hep-ph/9804321.

\bibitem{madgr}
 T.~Stelzer and W.~F.~Long, Comput. Phys. Commun. {\bf 81}, 357 (1994).

\bibitem{Yao} W.-M.~Yao, in {\it Proceedings of the 1996 DPF/DPB
Summer Study on New Directions for High-Energy Physics}, Snowmass,
edited by D.~Cassel, L.~Gennari, and R.~Siemann (SLAC, Menlo Park,
1997), p.~619.

\end{references}
\end{document}